\documentclass[aps,pra,pdf,superscriptaddress,twocolumn,showpacs,nofootinbib]{revtex4-2}
\usepackage{amsmath,amssymb,amsfonts}
\usepackage{eqnarray}
\usepackage{graphics,float}
\usepackage{bm}
\usepackage{microtype}
\usepackage{braket}
\usepackage{amsmath}
\usepackage{physics}
\usepackage{mathtools}
\usepackage{gensymb}
\usepackage{color,xcolor,colortbl}
\usepackage{tikz}
\usepackage{soul}
\usetikzlibrary{shapes}
\usepackage[colorlinks=true,
linkcolor=blue,
filecolor=magenta,      
urlcolor=blue,
citecolor=blue]{hyperref}
\usepackage{lipsum}
\usepackage{orcidlink}
\usepackage[normalem]{ulem}

\begin{document}

\title{Kelvin-Helmholtz and counter-superflow instabilities driven by differential effective rotation in a three-component Bose-Einstein condensate}

\author{Susovan Giri $^\S$ \orcidlink{0009-0003-3848-8662}}
\email{iamsusovangiri@gmail.com}
\affiliation{Department of Physics, Indian Institute of Technology Kharagpur, Kharagpur, West Bengal 721302, India}

\author{Arpana Saboo $^\S$ \orcidlink{0009-0000-8462-4547}}
\email{arpana.saboo@gmail.com}
\affiliation{Department of Physics, Indian Institute of Technology Kharagpur, Kharagpur, West Bengal 721302, India}
\affiliation{RIKEN Center for Emergent Matter Science (CEMS), Wako, Saitama 351-0198, Japan}

\author{Hari Sadhan Ghosh \orcidlink{0009-0003-9969-390X}}
\affiliation{Department of Physics, Indian Institute of Technology Kharagpur, Kharagpur, West Bengal 721302, India}

\author{Vipin \orcidlink{0009-0002-5841-062X}}
\affiliation{Department of Physics, Indian Institute of Technology Kharagpur, Kharagpur, West Bengal 721302, India}

\author{Sonjoy Majumder \orcidlink{0000-0001-9131-4520}}
\email{sonjoym@phy.iitkgp.ac.in}
\affiliation{Department of Physics, Indian Institute of Technology Kharagpur, Kharagpur, West Bengal 721302, India}

\begin{abstract}
Interfacial hydrodynamic instabilities in multicomponent superfluids provide a versatile platform to explore nonequilibrium quantum dynamics beyond classical fluid analogues. We study dynamical interfacial instabilities in a quasi-two-dimensional three-component Bose-Einstein condensate confined in a harmonic trap, where differential effective rotation between neighboring components is generated through selective phase imprinting. A component with an imprinted vorticity $l_j$ acquires an azimuthal phase winding $\phi_j = l_j\theta$, producing a superfluid velocity $v_\theta(r_j) = l_j\hbar/mr_j$ at interface radius $r_j$ and an effective angular velocity $\Omega_j^{\mathrm{eff}} = l_j\hbar/mr_j^2$. The differential effective rotation $\Delta\Omega_j^{\mathrm{eff}} = \Omega_j^{\mathrm{eff}} - \Omega_{j+1}^{\mathrm{eff}}$ between neighboring components constitutes the sole driver of interfacial shear and the Kelvin-Helmholtz instability (KHI). Three instability regimes are examined: KHI in the strongly immiscible limit, counter-superflow instability in the partially miscible regime, and a parameter window where both mechanisms coexist. A hydrodynamic pressure-balance model is derived, showing that the KHI onset condition depends only on $\Delta\Omega_j^{\mathrm{eff}}$. Time-dependent Gross-Pitaevskii simulations and a Bogoliubov-de Gennes analysis identify the dominant unstable modes, whose spatial structures are consistent with the observed density modulations. The dominant azimuthal mode is set by a competition between shear-driven destabilization and surface-tension stabilization, confirmed by numerical simulations with different imprinted vorticities. The results show that a three-component condensate with two tunable interfaces provides a controlled platform for studying multicomponent quantum hydrodynamics, with instability dynamics significantly richer than those of binary mixtures.
\end{abstract}

\maketitle
\def\thefootnote{\S}\footnotetext{These authors have contributed equally to this work.}\def\thefootnote{\arabic{footnote}}

\section{Introduction}
\label{sec:intro}

Superfluids, owing to the absence of viscosity, support coherent flow dynamics in which perturbations can evolve into structured density patterns and collective excitations. In Bose-Einstein condensates (BECs) with multiple interacting components, relative motion between species and the formation of distinct phase boundaries create favorable conditions for the emergence of hydrodynamic instabilities. Multi-component BECs provide an extended setting in which such instabilities may develop simultaneously at multiple interfaces or coexist with regions of partial miscibility, enabling intriguing interfacial and surface dynamics. Recent studies have demonstrated several phenomena intrinsic to this setting, including effective surfactant behavior~\cite{jimbo_surfactant_2021, pub.1028358591, PhysRevA.58.4836, PhysRevA.66.013612, PhysRevA.65.063614, PhysRevA.66.015602, PhysRevA.65.033618, PhysRevA.67.053608}, Rayleigh-Taylor instability~\cite{sasaki_rayleigh-taylor_2009,gautam_rayleigh-taylor_2010,burmistrov_rayleigh-taylor_2009, saboo_rayleigh-taylor_2023, y226-17w9, PhysRevA.85.013630, doi:10.1126/sciadv.adw9752, PhysRevA.83.043623, PhysRevA.85.013602, PhysRevA.80.033609, PhysRevA.79.023606}, Rayleigh-Plateau instability~\cite{PhysRevA.107.063312}, Richtmyer-Meshkov instability~\cite{bezett_magnetic_2010, PhysRevA.83.033602}, Faraday instability~\cite{maity_parametrically_2020,kwon_spontaneous_2021, tang_faraday_2011, PhysRevA.105.063319, PhysRevA.70.011601, PhysRevA.108.063317, PhysRevA.108.053315, PhysRevA.85.023613, PhysRevA.76.063609, abdullaev_faraday_2019, PhysRevA.89.023609, abdullaev_faraday_2015, sudharsan_faraday_2016, hernandez-rajkov_faraday_2021, PhysRevA.83.013603, chd2-slgy, PhysRevA.86.023620, PhysRevLett.89.210406, PhysRevX.9.011052, PhysRevLett.128.210401}, fingering instability~\cite{PhysRevA.97.023625}, capillary instability~\cite{sasaki_capillary_2011}, Crow instability~\cite{PhysRevA.84.021603, prasad_crow_2025, berloff_motion_2001}, modulational instability~\cite{mithun_modulational_2012,PhysRevA.97.011604, PhysRevA.96.041601, PhysRevA.65.021602, PhysRevLett.92.040401}, snake instability~\cite{PhysRevA.93.033618,PhysRevA.88.043639,PhysRevA.110.L061303,PhysRevA.65.043612,2wc6-pp3r}, Rosensweig instability~\cite{PhysRevLett.102.230403}, and wetting phase transitions~\cite{indekeu_three-component_2025, PhysRevLett.93.210402, PhysRevA.91.013626}, etc. Despite this progress, dynamical instabilities in condensates with more than two components remain comparatively underexplored, motivating further investigation. Among the dynamical instabilities that arise from relative motion in multicomponent condensates, the hydrodynamic shear and counterflow mechanisms play a central role. In immiscible configurations, where components are separated by sharp interfaces, differential flow along the interface can trigger the Kelvin-Helmholtz instability (KHI)~\cite{PhysRevB.99.054104, 10.21468/SciPostPhys.17.3.076, takeuchi_quantum_2010,suzuki_crossover_2010,baggaley_kelvin-helmholtz_2018,PhysRevA.104.023312}, characterized by the amplification of interfacial perturbations and the subsequent nucleation of vortices~\cite{takeuchi_quantum_2010,suzuki_crossover_2010,baggaley_kelvin-helmholtz_2018,PhysRevA.104.023312,PhysRevLett.89.200403,PhysRevX.7.021031,zyvv-jv7z}. The nonlinear development of KHI often leads to complex vortex dynamics and, in extreme cases, quantum turbulence~\cite{das_vortex_2022,PhysRevResearch.5.043081,PhysRevResearch.6.L042003}.

In contrast, when components are miscible and spatially overlapping, relative counterflow gives rise to counter-superflow instability (CSI), a mechanism unique to quantum fluids with multiple superfluid components~\cite{takeuchi_quantum_2010,suzuki_crossover_2010,baggaley_kelvin-helmholtz_2018,PhysRevA.104.023312,PhysRevA.83.063602,PhysRevLett.105.205301}. Unlike KHI, the CSI manifests itself in the bulk density, and its phase modulations extend throughout the overlapping region rather than being confined to the interface. The distinct physical origins and mode structures of these instabilities highlight their intriguing dynamical behavior accessible in multicomponent condensates, particularly in situations where miscible and immiscible regions may coexist. The combination of interspecies interactions and relative flow at the interfaces shapes a variety of dynamical responses, including quantum pressure~\cite{takeuchi_quantum_2010,suzuki_crossover_2010,baggaley_kelvin-helmholtz_2018,maity_parametrically_2020} and vortex-mediated excitations~\cite{PhysRevLett.89.200403,PhysRevX.7.021031}, underscoring the relevance of KHI and CSI in condensates with more than two components as a platform to explore nonlinear interfacial and surface dynamics.

In quantum fluids, the quantum pressure term~\cite{harko_bose-einstein_2015,PhysRevA.103.023322,PhysRevA.106.043309} acts as a regularizing mechanism at short scales, suppressing unphysical divergences, while enabling the formation of distinct nonlinear structures, such as vortices~\cite{takeuchi_quantum_2010,suzuki_crossover_2010,baggaley_kelvin-helmholtz_2018,PhysRevA.104.023312}, spikes~\cite{bezett_magnetic_2010}, and bubble-shaped~\cite{bezett_magnetic_2010,gautam_rayleigh-taylor_2010, PhysRevA.83.033602} deformations, driving small-scale hydrodynamic instabilities. These nonlinear effects have significant implications for understanding quantum turbulence~\cite{das_vortex_2022,PhysRevResearch.5.043081,PhysRevResearch.6.L042003}, astrophysical superfluid~\cite{zyvv-jv7z, Melatos_2007}, and the broader field of non-equilibrium quantum fluids.

\begin{figure}[t]
    \centering
    \includegraphics[width=1\linewidth]{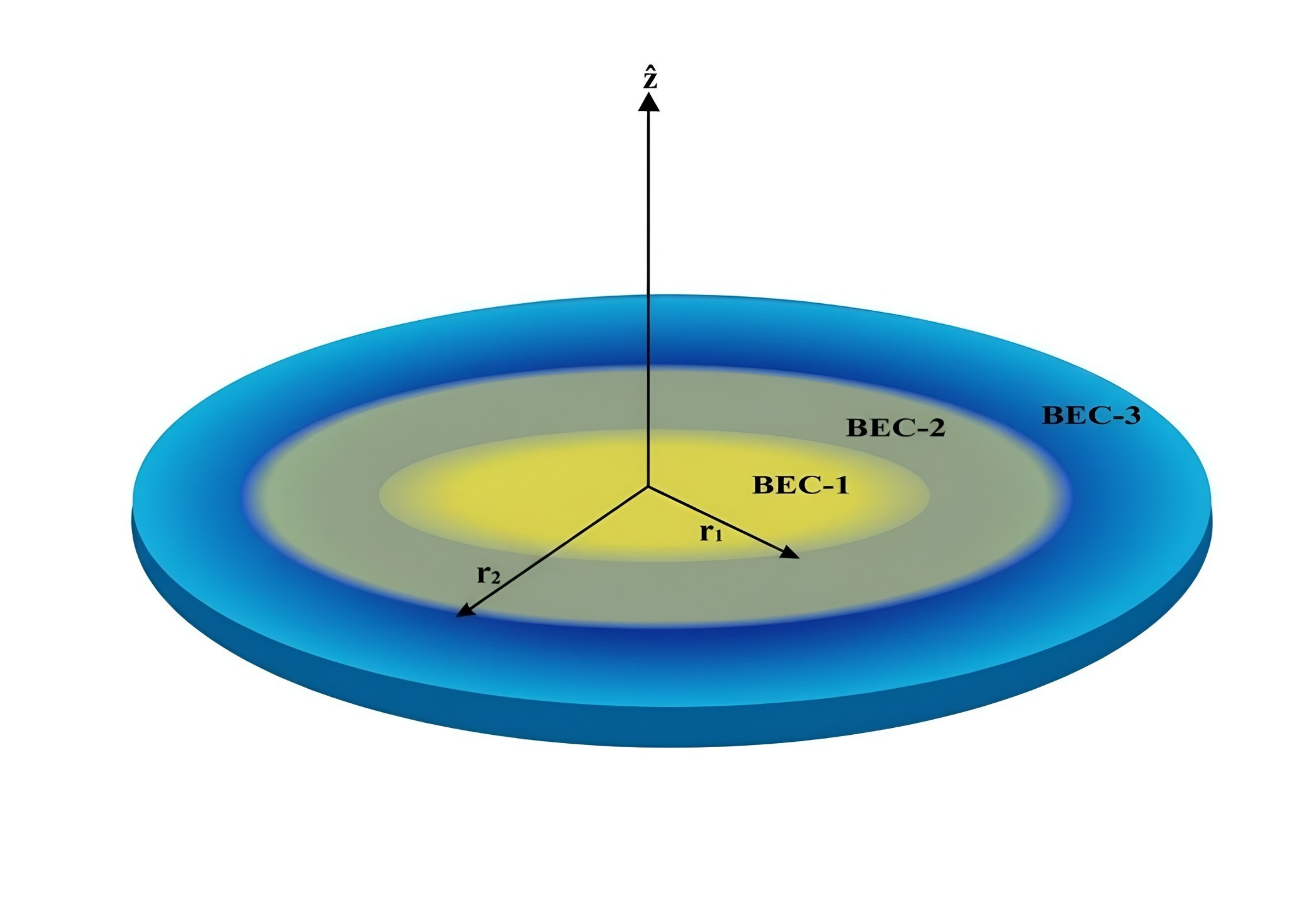}
    \caption{Schematic representation of a radially phase-separated three-component BEC in a quasi-2D harmonic trap.}
    \label{fig:1}
\end{figure}

In this work, we investigate the development of KHI and CSI in a three-component BEC confined in a quasi-two-dimensional (quasi-2D), axisymmetric harmonic trap. Strong intercomponent repulsion drives the system into a radially phase-separated configuration comprising three concentric shells with two well-defined circular interfaces, as illustrated in Fig.~\ref{fig:1}. Unlike binary mixtures, which possess a single interface, the three-component geometry admits a qualitatively richer landscape. The three-component radially stratified configuration has recently attracted attention as a platform for studying wetting transitions~\cite{indekeu_three-component_2025}, Rayleigh-Taylor instability~\cite{saboo_rayleigh-taylor_2023}, and vector rogue waves~\cite{bougas2025observationvectorroguewaves}, underscoring its experimental accessibility.

Interfacial velocity shear is introduced through selective phase imprinting on individual components, a protocol that generates an exact differential effective rotation between neighboring components without any rotating trap potential. Using hydrodynamic pressure-balance equations~\cite{suzuki_crossover_2010}, we analytically derive the KHI onset condition at each interface and demonstrate that it depends solely on the vorticity contrast between components. Time-dependent Gross-Pitaevskii  (GP) simulations reveal three distinct dynamical regimes: KHI in the strongly immiscible limit, CSI in the partially miscible regime, and a coexistence regime in which both instabilities are simultaneously triggered by a dynamical scattering-length quench, since the two interfaces can be placed in different miscibility regimes simultaneously, KHI and CSI develop concurrently at different spatial locations within the same condensate, a scenario with no binary analogue. A Bogoliubov-de Gennes (BdG) analysis further identifies the dominant unstable modes, showing that the preferred azimuthal mode number tracks the differential effective rotation at each interface rather than being fixed by the trap geometry. Notably, the intermediate component BEC-2, bounded by both interfaces simultaneously, inherits contributions from instabilities at each boundary and serves as a natural spectroscopic probe of the coupled interface dynamics; a feature unique to the three-component geometry. Together, these results establish the three-component BEC as a controlled and experimentally accessible platform for studying multicomponent quantum hydrodynamics with an instability richness that binary superfluids fundamentally cannot match.

The structure of the paper is as follows. Section~\ref{Theory} describes our theoretical framework. In Section~\ref{Numerical}, we discuss the numerical results. In Section~\ref{BdG}, we track the most dominant low-lying excitations obtained from the BdG analysis. In Section~\ref{Conclusions}, we summarize and conclude.

\section{Theory}
\label{Theory}

We consider a three-component BEC in which the second component (BEC-2) is sandwiched between the other two components, BEC-1 and BEC-3, as illustrated in Fig.~\ref{fig:1}. 
The dynamics of each component in a harmonic trap is described by the coupled GP equations~\cite{PhysRevA.65.023603,PhysRevA.84.053623,PhysRevA.72.043613,PhysRevA.80.063621,pethick_boseeinstein_2008}:
\begin{equation}
    i\hbar \frac{\partial \psi_j}{\partial t} = \left[ -\frac{\hbar^2}{2m_j} \nabla^2 + U_j + \sum_{j'=1}^3 g_{jj'} |\psi_{j'}|^2 \right] \psi_j .
\label{eqn::RNLS}    
\end{equation}

Here $\psi_j~(j = 1, 2, 3)$ is the macroscopic wave function of the $j$-th component with mass $m_j$. $U_j = \frac{1}{2} m_j \omega_r^2 \left( r^2 + \lambda^2 z^2 \right)$ is the circularly symmetric trapping potential that confirms the quasi-2D structure of the condensate with aspect ratio $\lambda = \omega_z / \omega_r \gg 1$, where \( \omega_r \) is the radial trapping frequency. The contact interactions are described by the coupling constants $g_{jj'} = 2\pi \hbar^2 a_{jj'} / m_{jj'}$, where $a_{jj'}$ denotes the $s$-wave scattering length and $m_{jj'} = m_j m_{j'} / (m_j + m_{j'})$ is the reduced mass between components $j$ and $j'$.

Differential effective rotation between neighboring components is introduced through selective phase imprinting~\cite{PhysRevA.97.043615}. A component with vorticity $l_j$ acquires an exact azimuthal phase winding $\phi_j = l_j\theta$. The corresponding superfluid azimuthal velocity at radius $r_j$ is:
\begin{equation}
v_{\theta,j}(r_j) = \frac{\hbar}{m_j}\frac{1}{r_j}\frac{\partial \phi_j}{\partial\theta} = \frac{l_j\hbar}{m_jr_j},
\label{eqn::velocity}
\end{equation}
giving an effective angular velocity at interface radius $r_j$:
\begin{equation}
\Omega_j^{\mathrm{eff}}(r_j) = \frac{v_{\theta,j}(r_j)}{r_j} = \frac{l_j\hbar}{m_jr_j^2}.
\label{eqn::Feynman}
\end{equation}
This follows exactly and directly from the phase $\phi_j = l_j\theta$. The differential effective rotation between neighboring components $j$ and $j+1$ at their shared interface $r_j$ is:
\begin{equation}
\Delta\Omega_j^{\mathrm{eff}} = \Omega_j^{\mathrm{eff}}(r_j) - \Omega_{j+1}^{\mathrm{eff}}(r_j) = \frac{(l_j - l_{j+1})\hbar}{m_jr_j^2},
\label{eqn::DeltaOmega}
\end{equation}
which is the physical driver of interfacial shear, as derived below. Note that $\Omega_{j+1}^{\mathrm{eff}}(r_j)$ denotes the effective angular velocity of component $j+1$ evaluated at the radius of interface $j$, not at any radius intrinsic to component $j+1$ itself; the same vorticity $l_{j+1}$ therefore yields a different numerical $\Omega^{\mathrm{eff}}$ at each of the two interfaces it borders.

The miscibility phase of any two components is governed by the parameter
$\Delta = g_{jj'}/\sqrt{g_{jj} g_{j'j'}} - 1$.
For $\Delta \ll 1$, the system is weakly immiscible, leading to partial
separation with a relatively broad interface. In contrast, for $\Delta > 1$,
the components are strongly segregated and the interface becomes sharp.
Thus, the interface thickness is directly controlled by the interaction
parameter $\Delta$~\cite{suzuki_crossover_2010,PhysRevA.66.013612,PhysRevLett.81.5718,PhysRevA.58.4836}. The stability of the radially stratified ground state (BEC-1 in the core, BEC-2 in the intermediate shell, BEC-3 in the exterior) rests on two conditions. First, each pair of neighboring components must satisfy the immiscibility condition $g_{jj'}^2 > g_{jj}g_{j'j'}$ ($\Delta > 0$), ensuring phase separation with a well-defined interface. Second, the radial ordering is set by the scattering-length hierarchy $a_{11} < a_{22} < a_{33}$: the component with the weakest self-interaction occupies the core, minimizing the total interaction energy~\cite{PhysRevA.80.033609, PhysRevA.79.023606}. This stratified configuration was established and verified in our earlier three-component Rayleigh-Taylor study using the same scattering parameters~\cite{saboo_rayleigh-taylor_2023}. We confirm numerically that imaginary-time propagation converges to this configuration from generic initial conditions, confirming it is the global energy minimum for the interaction parameters considered.

Within the Thomas-Fermi approximation, KHI appears in the
immiscible regime where the interatomic interaction satisfies $g_{jj'}^2 > g_{jj} g_{j'j'}$, ensuring negligible overlap between components. We assume sharp interfaces at the radii $r = r_1$ (between BEC-1 and BEC-2) and $r = r_2$ (between
BEC-2 and BEC-3), with interface tensions $\sigma_j$ ($j = 1, 2$).

To model the KHI hydrodynamically, we introduce a small angular perturbation at each interface defined by the radial variable $r_j$ ($j = 1, 2$), and the modified interface can be expressed as~\cite{suzuki_crossover_2010}:
\begin{equation}
\eta_j=r_j + \alpha_j \sin(k_\theta \theta - \omega t).
\label{eqn::Curve}
\end{equation}
Here, the perturbation amplitude $ \alpha_j\ll r_j$. $k_\theta$ and $\omega$ are the azimuthal mode number and modulation frequency of the perturbation, respectively.

In the hydrodynamic approximation, the wave function for the $j$-th component is written as: 
\begin{equation}
\psi_j = \sqrt{n_j(r)} \exp [ i\phi_j(r,\theta,t)].
\label{eqn::Stationary wavefunction}
\end{equation}
To model the excitation of surface waves, we introduce phase perturbations~\cite{suzuki_crossover_2010}:
\begin{equation}
\phi_j= l_j \theta + A_j^\pm e^{ \pm k_r r} \cos(k_{\theta} \theta - \omega t),
\label{eqn::Phase}
\end{equation}
where $l_j$ and $A_j$ are the vorticity and infinitesimal parameter of the j-th component. The sign depends on whether the condensate is inside ($+$) or outside ($-$) of the interface. For incompressible flow, $k_r$ and $k_\theta$ are related through Laplace's equation for the phase.

The radial velocity derived from the superfluid phase is $v_{r,j}=\hbar / m_j \partial_r \phi_j$. The kinematic boundary condition requires that, at the interface $r=\eta_j$, the radial velocity of the interface equals the radial velocity of the fluid~\cite{suzuki_crossover_2010}:
\begin{equation}
\partial_t \eta_j + \frac{l_j \hbar}{m_j r_j^2} \partial_\theta \eta_j=v_{r,j}=\hbar / m_j \partial_r \phi_j,
\label{eqn::Kinematic condition}
\end{equation}
where $l_j\hbar/m_jr_j^2 \equiv \Omega_j^{\mathrm{eff}}(r_j)$ is the exact azimuthal velocity of the central vortex at the interface [Eq.~\eqref{eqn::Feynman}]. This gives:
\begin{equation}
\pm  A_j^\pm e^{\pm k_r \eta_j} = \frac{m_j \alpha_j}{\hbar k_r}\left(\frac{l_j \hbar}{m_j r_j^2} k_\theta - \omega\right).
\label{eqn::Substitution}
\end{equation}

The dynamic condition at the interface follows from the Bernoulli equation. The hydrodynamic pressure for each component is:
\begin{equation}
P_j = - \hbar n_j  \left(\frac{\partial \phi_j}{\partial t} + \frac{l_j \hbar}{m_j r_j^2} \frac{\partial \phi_j}{\partial \theta}\right).
\label{eqn::Pressure}
\end{equation}
At $r = r_j$:
\begin{equation}
P_j = - \hbar n_j \left( \omega - \frac{l_j \hbar}{m_j r_j^2} k_\theta\right) A_j^\pm e^{ \pm k_r \eta_j} \sin(k_{\theta} \theta - \omega t).
\label{eqn::Pressure2}
\end{equation}

Therefore, the total pressure jump at the interface $j$ is:
\begin{equation}
\begin{split}
P_{j+1} - P_j = -\frac{\alpha_j \sin(k_\theta \theta - \omega t)}{k_r}
\Big[\rho_j\Big(\omega-\frac{l_j \hbar}{m_j r_j^2} k_\theta\Big)^2 \\
+\,\rho_{j+1}\Big(\omega-\frac{l_{j+1} \hbar}{m_{j+1} r_j^2} k_\theta\Big)^2\Big],
\end{split}
\label{eqn::Pressure3}
\end{equation}
where $\rho_j = m_jn_j (r_j )$ is 2D mass density at the $j$-th interface.

At the interface, the total pressure difference between
$(j+1)$- and $j$-components must balance the Laplace pressure arising
from surface tension:
\begin{equation}
    P_{j+1}- P_j= -\sigma_j \frac{k_\theta^2}{r_j^2}\alpha_j \sin(k_\theta \theta - \omega t) .
\label{pressure4}
\end{equation}

Using Eqs.~\eqref{eqn::Pressure3} and~\eqref{pressure4}, taking $m_j=m_{j+1}=m$, and canceling $\alpha_j \sin(k_\theta \theta - \omega t)$, we obtain a quadratic equation with respect to $\omega$:
\begin{equation}
\begin{aligned}
\frac{\rho_j}{k_r} \left(\omega - \frac{l_j \hbar}{m r_j^2} k_\theta\right)^2  + \frac{\rho_{j+1}}{k_r} \left(\omega - \frac{l_{j+1} \hbar}{m r_j^2} k_\theta\right)^2 =\frac{\sigma_j k_\theta^2}{r_j^2}.
\end{aligned}
\label{eqn::Angular velocity4}
\end{equation}
The solution of equation~\eqref{eqn::Angular velocity4} yields the dispersion relation~\cite{suzuki_crossover_2010}:
\begin{equation}
\begin{aligned}
\omega_j & = \frac{\left[\rho_j \frac{l_j \hbar}{m r_j^2} + \rho_{j+1} \frac{l_{j+1} \hbar}{m r_j^2}\right]k_\theta}{\rho_j + \rho_{j+1}} 
\\&\quad \pm \sqrt{-\frac{\rho_j \rho_{j+1} \left(\frac{l_j \hbar}{m r_j^2} - \frac{l_{j+1} \hbar}{m r_j^2}\right)^2 k_\theta^2}{(\rho_j + \rho_{j+1})^2} + \frac{\sigma_j k_r k_\theta^2}{r_j^2(\rho_j + \rho_{j+1})}.}
\end{aligned}
\label{eqn::Frequency1}
\end{equation}
Instability occurs when the square root becomes imaginary. Using Eq.~\eqref{eqn::DeltaOmega}, the KHI onset condition is written as:
\begin{equation}
\left(\Delta\Omega_j^{\mathrm{eff}}\right)^2 > \frac{\sigma_j k_r}{r_j^2}\cdot \frac{(\rho_j + \rho_{j+1})}{\rho_j \rho_{j+1}},
\label{eqn::KHI_unified}
\end{equation}
the superfluid analogue of the classical KHI velocity-shear condition. The instability is driven solely by the vorticity contrast $\Delta l_j = l_j - l_{j+1}$ between neighboring components. For a system with only BEC-2 carrying vorticity $l_2$ ($l_1 = l_3 = 0$):
\begin{equation}
\text{Inner interface } (r_1): \quad 
l_2^2 > \frac{m^2 r_1^2 \sigma_1 k_r}{\hbar^2} \cdot \frac{\rho_1 + \rho_2}{\rho_1 \rho_2} .
\label{eqn::Critical Frequency1}
\end{equation}
\begin{equation}
\text{Outer interface } (r_2): \quad 
l_2^2 > \frac{m^2 r_2^2 \sigma_2 k_r}{\hbar^2} \cdot \frac{\rho_2 + \rho_3}{\rho_2 \rho_3} .
\label{eqn::Critical Frequency2}
\end{equation}

In contrast, CSI arises in the miscible regime, where
$g_{jj'}^2 < g_{jj} g_{j'j'}$~\cite{y226-17w9, PhysRevA.94.013602, PhysRevA.80.033609}.
CSI does not rely on a sharp interface; instead, it originates from relative counterflow within overlapping components. When the vorticity exceeds a critical value determined by the interaction strengths and densities, density modulations develop across the overlap region. The hydrodynamic description must therefore account for compressible excitations rather than purely interfacial waves.

\section{NUMERICAL RESULTS}
\label{Numerical}
In our numerical demonstration of KHI and CSI for a three-component BEC, we use $^{87}$Rb atoms with their three hyperfine states.
State $|F = 2, m_F = 0\rangle$ (BEC-2) is positioned between the $|F = 1, m_F = -1\rangle$ (BEC-1) and $|F = 1, m_F = 0\rangle$ (BEC-3) states~\cite{bougas2025observationvectorroguewaves, saboo_rayleigh-taylor_2023}, as described in Fig.~\ref{fig:1}, trapped  
in a quasi-2D harmonic potential.
We choose \(^{87}\text{Rb}\)-condensate due to their well-known tunability of atom-atom interactions \cite{PhysRevLett.89.283202} and hyperfine-state selective responses~\cite{PhysRevA.94.013427}. The split-step Crank-Nicolson method~\cite{Crank_Nicolson_1947,PhysRevE.65.016703, Muruganandam_2009, PhysRevE.95.023310, LE2024160, VUDRAGOVIC20122021} is employed in imaginary time propagation to numerically obtain the ground state of a phase-separated three-component BEC in the absence of vorticity ($l_j=0$). No rotation is applied at any stage of the numerical protocol. 

To study the onset of instability and its subsequent evolution, we perform real-time propagation by solving Eq.~\eqref{eqn::RNLS} starting from the ground state with an imprinted phase. In our simulations, the spatial domain spans \( -20l \) to \( 20l \) along \( x \) and \( y \) using a grid \( 801 \times 801 \). Here, \( l = \sqrt{\hbar/m_j \omega_r}= 1.52433 \times 10^{-6} \) m is the characteristic length, where $m_j=1.45 \times 10^{-25}$ kg. The dynamics of the system is studied with resolution  \( \Delta x = \Delta y = 0.05l \) and a time step of \( \delta t = (2 \times 10^{-4})/\omega_r \).
The radial trapping frequency \(\omega_r = 2\pi \times 50\,\text{Hz}\) and an aspect ratio of \(\lambda = 50\).

We present the numerical results in the following three cases: KHI in Section~\ref{KHI}, CSI in Section~\ref{CSI}, Coexistence regime in Section~\ref{KHICSI}. In all cases, the differential effective rotation $\Delta\Omega_j^{\mathrm{eff}}$ at each interface is realized by phase imprinting of vorticity $l_j$ on the relevant component, giving an exact azimuthal phase $\phi_j = l_j\theta$ and an exact effective angular velocity $\Omega_j^{\mathrm{eff}} = l_j\hbar/m_jr_j^2$ at the interface [Eq.~\eqref{eqn::Feynman}]. 
Component-selective phase imprinting is achieved through internal-state addressing, exploiting the distinct hyperfine states of the three components~\cite{PhysRevA.94.013427}; selectivity is determined by the transition linewidth and detuning, not by beam waist or condensate size.

\subsection{Case I: Kelvin-Helmholtz Instability}
\label{KHI}

\begin{figure}[b]
	\centering
	\includegraphics[width=1\linewidth]{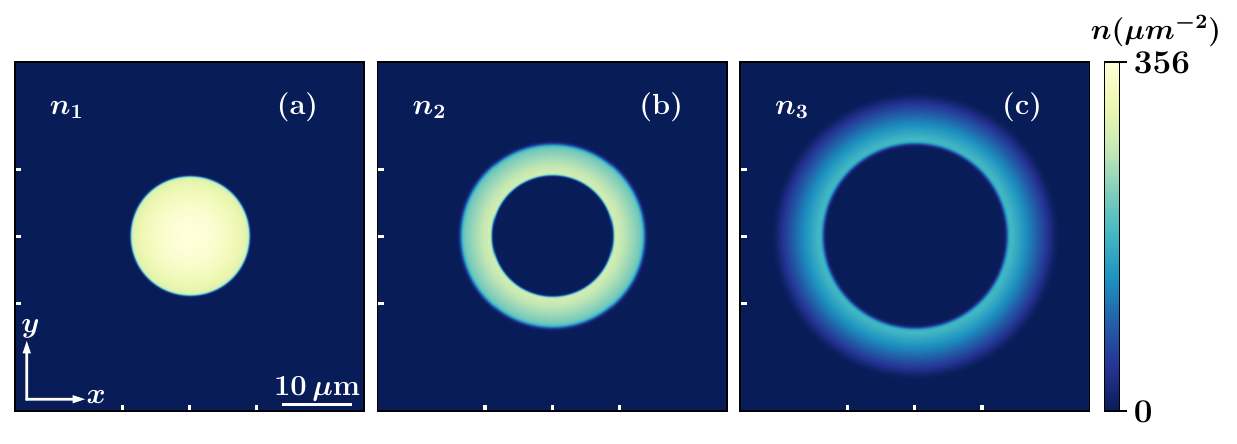}
	\caption{ Ground state density profile of the three components in a phase-separated BEC. The system is confined in a harmonic trap with $(\omega_r, \omega_z) = 2\pi \times (50,2500)$ Hz, and each component contains $N_j = 80,000$ atoms. The intracomponent scattering lengths are $a_{11}=92.4a_0$, $a_{22}=94.5a_0$, and $a_{33}=100.4a_0$, while the intercomponent scattering lengths are $a_{12}=a_{21}=213a_0$, $a_{13}=a_{31}=213a_0$, and $a_{23}=a_{32}=127a_0$, where $a_0$ denotes the Bohr radius.}
	\label{fig:2}
\end{figure}

To study KHI, completely phase separated three-component BEC is considered with the intraspecies and interspecies \(s\)-wave scattering lengths~\cite{saboo_rayleigh-taylor_2023} as \(a_{11} = 92.4a_0\), \(a_{22} = 94.5a_0\), \(a_{33} = 100.4a_0\), \(a_{12} = a_{21} = 213a_0\), \(a_{13} = a_{31} = 213a_0\), and \(a_{23} = a_{32} = 127a_0\), where \(a_0\) is the Bohr radius. The choice of $a_{11}<a_{22}<a_{33}$ along with the intercomponent interaction leads to a stable radially phase separated structure with BEC-1 in the core, BEC-2 in the intermediate shell and BEC-3 in the outer region, as shown in Fig.~\ref{fig:2}. The calculated interface thickness characteristic parameter $\Delta$ is $1.2$ and $0.3$ for the first and second interfaces, respectively~\cite{suzuki_crossover_2010,PhysRevA.66.013612,PhysRevLett.81.5718,PhysRevA.58.4836}. Their positive values indicate that the system is susceptible to KHI.

In the real-time dynamics of this prepared condensate, interfacial velocity shear is generated by imprinting a phase of vorticity $l_2$ on BEC-2 ($l_1 = l_3 = 0$), giving exact differential effective rotation $\Delta\Omega_j^{\mathrm{eff}} = l_2\hbar/mr_j^2$ at both interfaces. The interfacial tensions are $\sigma_{1} = 6.73 \times 10^{-16}$ N/m and $\sigma_{2} = 1.82 \times 10^{-16}$ N/m, calculated from Refs.~\cite{suzuki_crossover_2010,PhysRevA.66.013612,PhysRevA.78.023624,maity_parametrically_2020}. The analytical critical vorticity from Eqs.~\eqref{eqn::Critical Frequency1} and~\eqref{eqn::Critical Frequency2} using numerically computed values of $r_1$, $r_2$, $\rho_1$, $\rho_2$, $\rho_3$ gives $l_2 \approx 23$ and $l_2 \approx 19$ for the inner and outer interfaces respectively. Numerical simulations indicate the instability onset occurs at $l_2 \approx 8$ for both interfaces, in reasonable agreement with the analytical estimate given the sharp-interface Thomas-Fermi approximation. A vorticity $l_2 = 20$, above the numerical threshold, is used to trigger well-developed KHI within the simulation window, while $l_1 = l_3 = 0$.

\begin{figure}[t]
    \centering
    \includegraphics[width=1\linewidth]{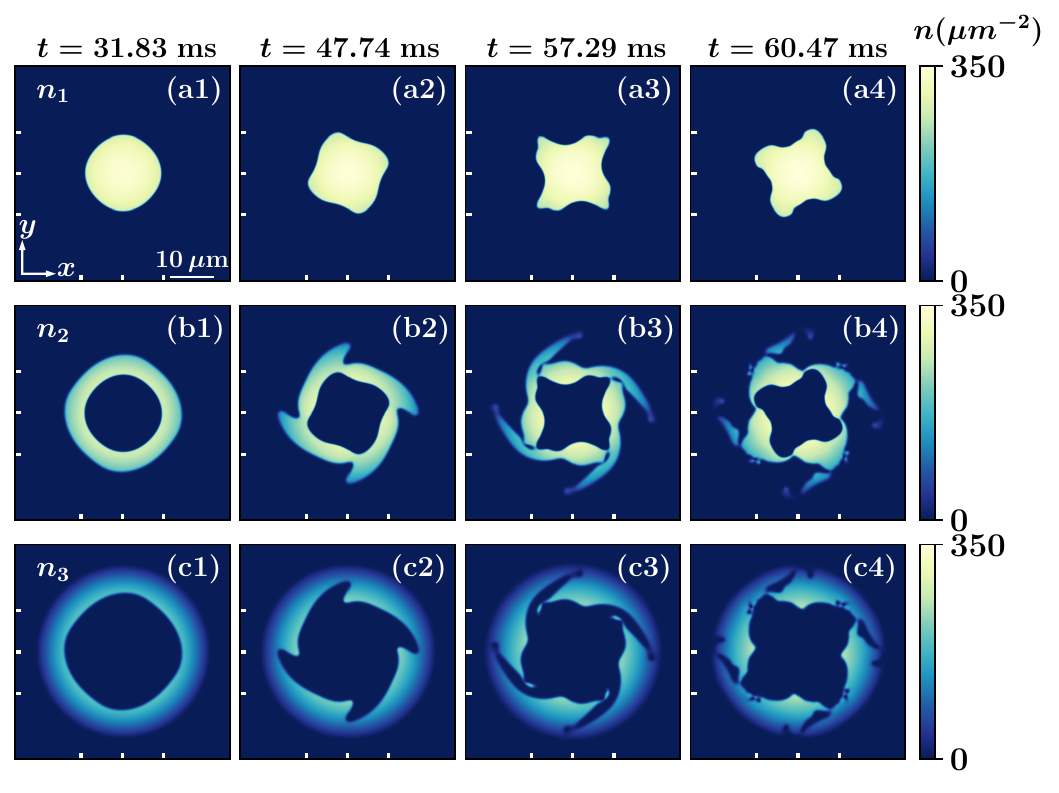}
    \caption{Real time evolution of the density distributions of a three-component BEC confined in quasi-2D harmonic trap for BEC-1, BEC-2 and BEC-3 in panels [(a1)-(a4)], [(b1)-(b4)], [(c1)-(c4)] respectively. A phase of vorticity $l_2 = 20$ is imprinted on BEC-2 ($l_1 = l_3 = 0$), realizing exact differential effective rotation $\Delta\Omega^{\mathrm{eff}} = l_2\hbar/mr_j^2$ at both interfaces. No rotation is applied. Snapshots at $t = 31.83$ ms, 47.74 ms, 57.29 ms, and 60.47 ms show the interfacial deformation and vortex nucleation accompanying KHI development. A full movie is available in the Supplemental Material \cite{supple}.}
    \label{fig:3}
\end{figure}

The density plots in Fig.~\ref{fig:3} show the real-time evolution of the three components. At $t = 31.83$ ms, the interfaces remain nearly circular though small distortions begin to appear, indicating the initial axisymmetry is broken by the imprinted vorticity. At $t = 47.74$ ms, clear four-fold wavy patterns develop at the second interface. The instability first develops at the interface between BEC-2 and BEC-3 because the interspecies interaction is weaker ($a_{23}<a_{12}$), leading to a lower interfacial tension $(\sigma_2<\sigma_1)$, and the outer interface experiences a larger effective centrifugal force due to the higher $\Delta\Omega_2^{\mathrm{eff}}$ at that interface. Over time at $t = 57.29$ ms the pattern intensifies and quantized vortices are nucleated around the kinks and troughs~\cite{takeuchi_quantum_2010, suzuki_crossover_2010}. At $t = 60.47$ ms all interfaces exhibit strong deformations and numerous quantized vortices indicate full KHI development.

\subsection{Case II: Counter-Superflow Instability}
\label{CSI}

\begin{figure}[t]
	\centering
	\includegraphics[width=1\linewidth]{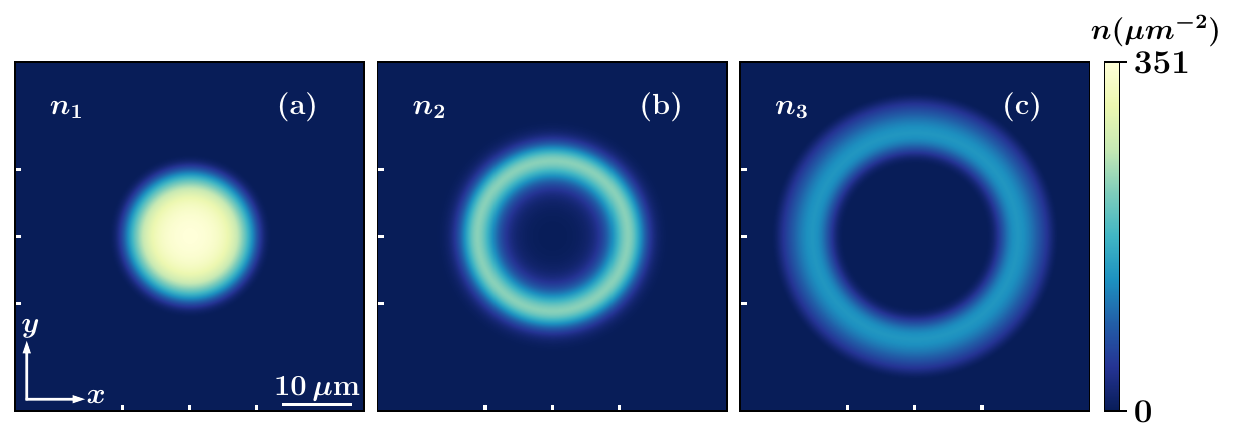}
	\caption{ Ground state density profile of a weakly miscible three-component BEC confined in quasi-2D harmonic trap. Trap frequencies are $(\omega_r, \omega_z) = 2\pi \times (50,2500)$ Hz, with $N_j = 80,000$ atoms in each component. The intracomponent scattering lengths are $a_{11}=92.4a_0$, $a_{22}=94.5a_0$, and $a_{33}=100.4a_0$, while the intercomponent scattering lengths are $a_{12}=a_{21}=93.4a_0$, $a_{13}=a_{31}=120a_0$, and $a_{23}=a_{32}=97a_0$, where $a_0$ denotes the Bohr radius.}
	\label{fig:4}
\end{figure}

CSI is a distinct non-linear phenomenon in multi-component quantum fluids that requires special configurations of atom-atom interactions among the components.
Compare to KHI case, here the intermediate component (BEC-2) exhibits partial miscibility with both BEC-1 and BEC-3, which are mutually immiscible to each other. To create this ground state as shown in Fig.~\ref{fig:4}, we set the interspecies and intraspecies scattering lengths to satisfy \( g_{11} g_{22} \approx g_{12}^2 \) and \( g_{22} g_{33} \approx g_{23}^2 \). The scattering parameters chosen for the three-component BEC are 
\( a_{11} = 92.4a_0 \), \( a_{22} = 94.5a_0 \), \( a_{33} = 100.4a_0 \),  \( a_{12} = a_{21} = 93.4a_0 \), \( a_{13} = a_{31} = 120a_0 \),  \( a_{23} = a_{32} = 97a_0 \). For this case, the negativeness of the $\Delta$ parameter favours the CSI criteria  and its calculated values are $-4.7 \times 10^{-4}$ and $-4.1 \times 10^{-3}$ for the first and second interfaces.

To study the morphological changes at the interfaces due to this counterflow instability, we imprint phase of vorticity $l_2 = 34$ on BEC-2 ($l_1 = l_3 = 0$). This creates differential effective angular velocity $\Omega_2^{\mathrm{eff}}(r_1) = l_2\hbar/mr_1^2$ at the inner interface and $\Omega_2^{\mathrm{eff}}(r_2) = l_2\hbar/mr_2^2$ at the outer interface, generating relative counterflow across both weakly miscible boundaries and setting up the conditions for CSI.

\begin{figure}[b]
    \centering
    \includegraphics[width=1\linewidth]{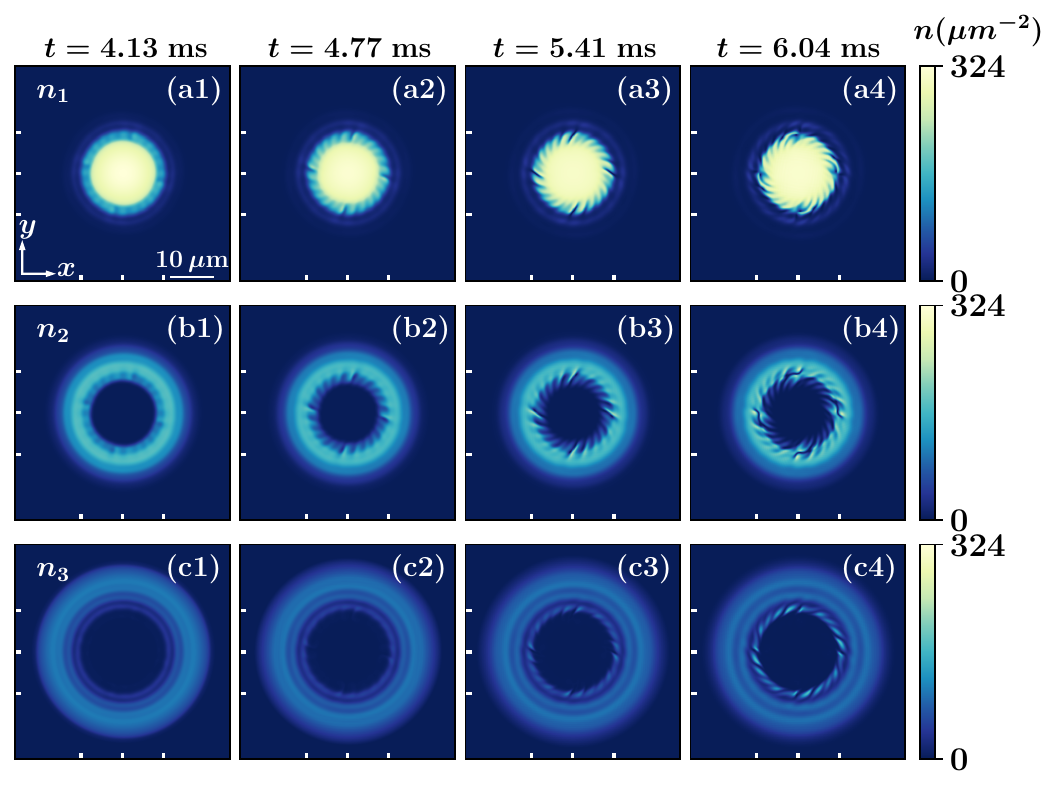}
    \caption{Real time evolution of the density distributions of a three-component BEC for BEC-1, BEC-2 and BEC-3 in panels [(a1)-(a4)], [(b1)-(b4)], [(c1)-(c4)] respectively, following phase imprinting $l_2 = 34$ on BEC-2 ($l_1 = l_3 = 0$). No rotation is applied. Snapshots at $t = 4.13$ ms, 4.77 ms, 5.41 ms, and 6.04 ms describe the interfacial deformation driven by CSI. A full movie is available in the Supplemental Material \cite{supple}.}
    \label{fig:5}
\end{figure}

Fig.~\ref{fig:5} shows the real-time evolution of the system. The first panel at $t = 4.13$ ms reveals initial deformations in the density profiles, with saw-tooth formation around the first and third components indicating the onset of broken axisymmetry. By $t = 4.77$ ms these modulations increase in amplitude along the interface [Figs.~\ref{fig:5}(a2-c2)]. At $t = 5.41$ ms the notches become very sharp at the interface between BEC-1 and BEC-2. At $t = 6.04$ ms both interfaces exhibit strong undulations. These features confirm that the observed symmetry breaking is driven by CSI. Unlike KHI, a vortex-antivortex pair rather than a single quantized vortex is nucleated across the interfacial boundary~\cite{suzuki_crossover_2010}.

\subsection{Case-III: Co-existence of Kelvin-Helmholtz and Counter-Superflow Instabilities}
\label{KHICSI}

\begin{figure}[b]
    \centering
    \includegraphics[width=1\linewidth]{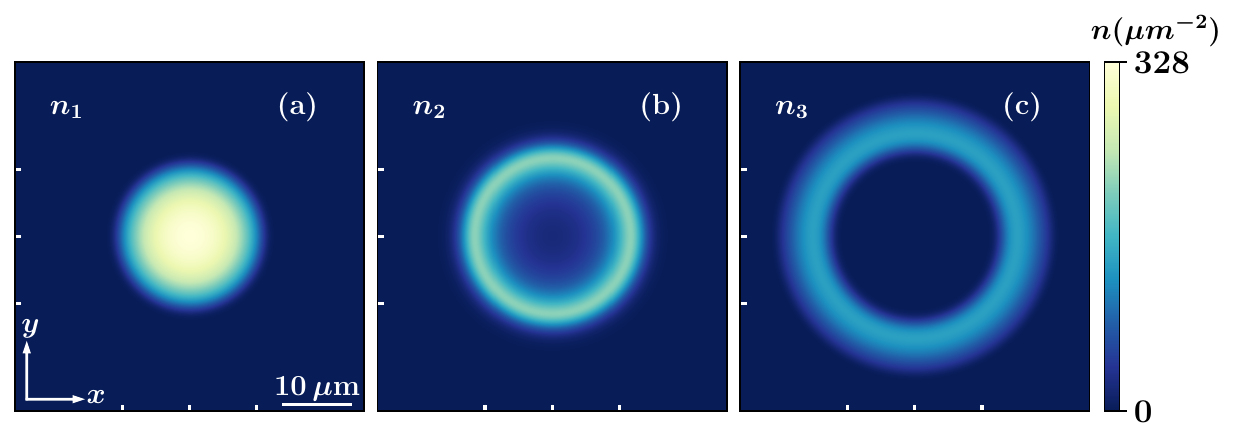}
    \caption{Ground state density profile of a weakly miscible three-component BEC confined in quasi-2D harmonic trap. Trap frequencies are $(\omega_r, \omega_z) = 2\pi \times (50,2500)$ Hz, with $N_j = 80,000$ atoms in each component. The intracomponent scattering lengths are $a_{11}=92.4a_0$, $a_{22}=94.5a_0$, and $a_{33}=100.4a_0$, while the intercomponent scattering lengths are $a_{12}=a_{21}=93.1a_0$, $a_{13}=a_{31}=120a_0$, and $a_{23}=a_{32}=97a_0$, where $a_0$ denotes the Bohr radius.}
    \label{fig:6}
\end{figure}

The simultaneous formation of KHI and CSI is a distinct superfluid phenomenon. The objective here is to induce both instabilities in a dynamically quenched condensate by tuning the atom-atom interaction to transform the system from a miscible phase to an immiscible phase while vorticity is already imprinted.
The first step here is to prepare the initial condensate configuration with fully immiscible BEC-1 and BEC-3, and BEC-2 exhibits partial miscibility with other components. Here, we differ the interspecies scattering length parameters from those in Case-II by setting $a_{12}=a_{21}=93.1a_0$ and prepare the ground state with the remaining parameters as in Case-II (see Fig.~\ref{fig:6}). 

To stimulate the desired dynamical phase transition, we imprint a phase of vorticity $l_2 = 18$ on BEC-2 ($l_1 = l_3 = 0$) and simultaneously quench the intercomponent scattering lengths linearly over 15 ms: $a_{12}=a_{21}$ increases from $93.1a_0$ to $96a_0$, while $a_{23}=a_{32}$ increases from $97a_0$ to $102a_0$. The interface thickness parameter $\Delta$ varies from $-3.6 \times 10^{-3}$ to $2.7 \times 10^{-2}$ and $-4.1 \times 10^{-3}$ to $4.7 \times 10^{-2}$ for the inner and outer interfaces respectively, driving the system from a CSI-favorable ($\Delta < 0$) to a KHI-favorable ($\Delta > 0$) state. Here, the the scattering-length quench drives the system from a CSI-favorable to a KHI-favorable regime, while differential effective rotation $\Delta\Omega^{\mathrm{eff}}$ between components triggers the instabilities.
 
\begin{figure}[t]
    \centering
    \includegraphics[width=1\linewidth]{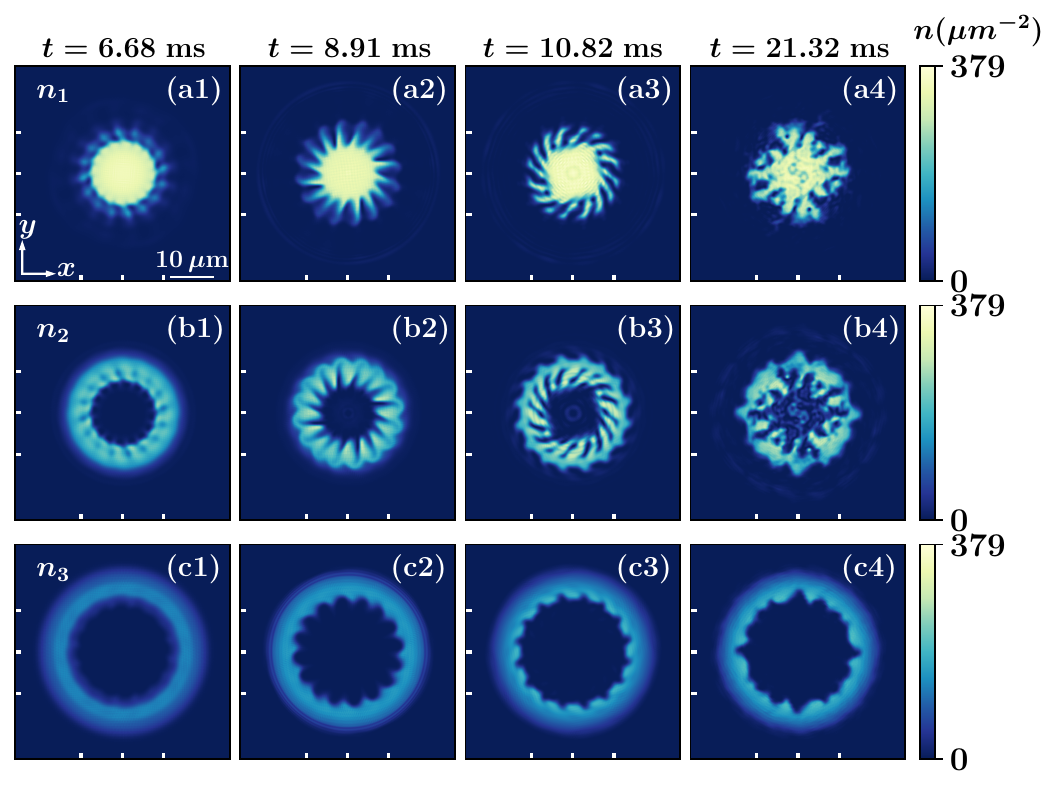}
    \caption{Real time evolution of the density distributions of a three-component BEC confined in quasi-2D harmonic trap are shown in panels [(a1)-(a4)], [(b1)-(b4)], and [(c1)-(c4)] for BEC-1, BEC-2 and BEC-3, respectively, following phase imprinting of vorticity $l_2 = 18$ on BEC-2 ($l_1 = l_3 = 0$) and simultaneous scattering-length quench. No rotation is applied. Snapshots at $t = 6.68$ ms, 8.91 ms, 10.82 ms, 21.32 ms describe the coexistence of KHI and CSI. A full movie of the dynamics is available in the Supplemental Material \cite{supple}.}
    \label{fig:7}
\end{figure}

At $t = 6.68$ ms, serrated structures emerge at both interfaces signaling the onset of CSI. At $t = 8.91$ ms these modulations amplify. At $t = 10.82$ ms four-fold patterns emerge at the first interface indicating the commencement of KHI [see Figs.~\ref{fig:7}(a3-c3)]. At $t = 21.32$ ms the interfaces become strongly corrugated and fragmented, demonstrating full development of both instabilities [see Figs.~\ref{fig:7}(a4-c4)].

\section{STABILITY ANALYSIS}
\label{BdG}

The dynamical instabilities observed in our three-component condensate originate from the amplification of specific collective excitation modes. While time-dependent GP simulations capture the nonlinear evolution of interfacial deformations and vortex nucleation, they do not directly reveal the microscopic nature of the unstable excitations that seed these patterns. A linear stability analysis is therefore essential to identify which modes become dynamically unstable and how their spatial structures distinguish different instability mechanisms. In particular, KHI is associated with surface modes localized near sharp interfaces, whereas CSI involves modes that extend into the overlapping bulk region. In a two-interface system with selective phase imprinting in different components, it is crucial to determine how these unstable modes are distributed spatially and which families dominate in different interaction regimes. To address this, we employ the BdG formalism to systematically characterize the low-lying excitations and identify the dominant unstable modes responsible for the observed dynamics. In the literature~\cite{takeuchi_quantum_2010,suzuki_crossover_2010}, calculations based on the BdG formalism identified the unstable excitation modes and delineated the conditions for the onset of KHI in two-component BEC, including the role of inter-particle interactions. However, it will be intriguing to study the same in three-component BEC in terms of correlation of similar or different instabilities and their mode-structures between different interfaces.
The interfacial instabilities observed here are numerically analyzed, and their excitation modes are estimated precisely using the BdG model.
We formulate a linear stability analysis for the three-component BEC system and identify low-lying collective modes and their excitation frequencies. This study reveals the most dominant unstable modes of KHI and CSI described in Section~\ref{Numerical}.

We begin by considering small perturbations around the stationary ground state of the $j$-th component described by

\begin{equation}
    \psi_j(\mathbf{r}, t) = e^{-i \mu_j t / \hbar} \left[ \psi_j(\mathbf{r},0) + \delta \psi_j(\mathbf{r}, t) \right],
\end{equation}
where $\psi_j(\mathbf{r},0)$ is the stationary state before the phase imprinting, and $\delta \psi_j(\mathbf{r}, t)$ is its collective excitation. We can write the excitation wavefunctions in the standard Bogoliubov form~\cite{PhysRevA.64.061603,PhysRevA.67.033602,PhysRevLett.86.564,PhysRevLett.78.1842} as

\begin{equation} 
\delta \psi_j (\mathbf{r}, t) = u_j(\mathbf{r}) e^{-i \omega t} + v_j^{*}(\mathbf{r}) e^{i \omega^{*} t}.
\end{equation}

We linearize the GP equation~\eqref{eqn::RNLS} around the stationary state with respect to $\delta \psi_j(\mathbf{r}, t)$ and get the BdG equations of the form $\hat{\mathcal{H}} \mathbf{M} = \hbar \omega \mathbf{M}$. The matrix $\hat{\mathcal{H}}$ is given by~\cite{SADAKA2024108948}

\begin{widetext}

\[
\hat{\mathcal{H}} =
\begin{pmatrix}
\hat{H}_1 &
g_{12}\psi_1\psi_2^{*} &
g_{13}\psi_1\psi_3^{*} &
g_{11}\psi_1^{2} &
g_{12}\psi_1\psi_2 &
g_{13}\psi_1\psi_3 
\\[6pt]
g_{21}\psi_2\psi_1^{*} &
\hat{H}_2 &
g_{23}\psi_2\psi_3^{*} &
g_{21}\psi_2\psi_1 &
g_{22}\psi_2^{2} &
g_{23}\psi_2\psi_3 
\\[6pt]
g_{31}\psi_3\psi_1^{*} &
g_{32}\psi_3\psi_2^{*} &
\hat{H}_3 &
g_{31}\psi_3\psi_1 &
g_{32}\psi_3\psi_2 &
g_{33}\psi_3^{2} 
\\[6pt]
-g_{11}{\psi_1^{*}}^{2} &
-g_{12}\psi_1^{*}\psi_2^{*} &
-g_{13}\psi_1^{*}\psi_3^{*} &
-\hat{H}_1 &
-g_{12}\psi_1^{*}\psi_2 &
-g_{13}\psi_1^{*}\psi_3 
\\[6pt]
-g_{21}\psi_2^{*}\psi_1^{*} &
-g_{22}{\psi_2^{*}}^{2} &
-g_{23}\psi_2^{*}\psi_3^{*} &
-g_{21}\psi_2^{*}\psi_1 &
-\hat{H}_2 &
-g_{23}\psi_2^{*}\psi_3 
\\[6pt]
-g_{31}\psi_3^{*}\psi_1^{*} &
-g_{32}\psi_3^{*}\psi_2^{*} &
-g_{33}{\psi_3^{*}}^{2} &
-g_{31}\psi_3^{*}\psi_1 &
-g_{32}\psi_3^{*}\psi_2 &
-\hat{H}_3
\end{pmatrix}
\]

with
\[
\hat{H}_j
= -\frac{\hbar^{2}}{2m}\nabla^{2}
+ U_j - \mu_j
+ 2 g_{jj} |\psi_{j}|^{2}
+ \sum_{j' \neq j} g_{jj'} |\psi_{j'}|^{2}
 ,
\]

\end{widetext}

where $\mathbf{M}= (u_1, u_2, u_3,\, v_1, v_2, v_3)^{T}$ represents the excitation vector.

We diagonalize the matrix $\hat{\mathcal{H}}$ to get the excitation eigenmodes of this system. Here, our aim is to study the contributions of different eigen modes to the time-evolved BEC states. We define $\mathrm{f}_j = u_j + v_j$ as the Bogoliubov mode functions~\cite{6d1g-671p,gqlr-bgj8}, which correspond to the density fluctuations of the condensate ground state, expressed as ${\delta n_j}^{\mathrm{gs}} = \psi_j(\mathbf{r},0) \mathrm{f_j}$. Each Bogoliubov mode is identified with an integer azimuthal quantum number $\mathrm{L}$, which specifies the angular symmetry of the corresponding density fluctuations in the condensate~\cite{suzuki_crossover_2010,6d1g-671p,gqlr-bgj8}. To understand the temporal behavior, we evaluate the density fluctuation at each instant as

\begin{equation}
\delta n_j(\mathbf{r}, t) = |\psi_j(\mathbf{r}, t)|^{2} - |\psi_j(\mathbf{r}, 0)|^{2}.
\end{equation}

To estimate the contribution of individual Bogoliubov modes of the dynamically evolved system, we compute the overlap between the  $\delta n_j(\mathbf{r},t)$  and $\delta n^{gs}_j$, defined as~\cite{6d1g-671p},
\begin{equation}
O_j(t) = 
\frac{\displaystyle \int \delta n_j(\mathbf{r}, t) \psi_j(\mathbf{r}, 0) \mathrm{f}_j \, d^{2}\mathbf{r}}
{\sqrt{\displaystyle \int \delta n_j^{2}(\mathbf{r}, t) \, d^{2}\mathbf{r}
       \int \{\psi_j(\mathbf{r}, 0)\}^2 \mathrm{f}_j^{2} \, d^{2}\mathbf{r}}}.
\end{equation}
The spectral response of each mode is then obtained by taking the Fourier transform of the overlap function,
\begin{equation}
A_j(\omega) = \int O_j(t) \, e^{i \omega t} \, dt.
\end{equation}

The maximum amplitude,
$\mathrm{A_{p}} = \max |A_j(\omega)|$ for $j$-th component of the condensate
serves as a quantitative measure of the strength of the mode in the evolution of the $j$-th component, while the corresponding frequency \( \omega \) characterizes its excitation energy.

\begin{figure}[t]
    \centering
    \includegraphics[width=1\linewidth]{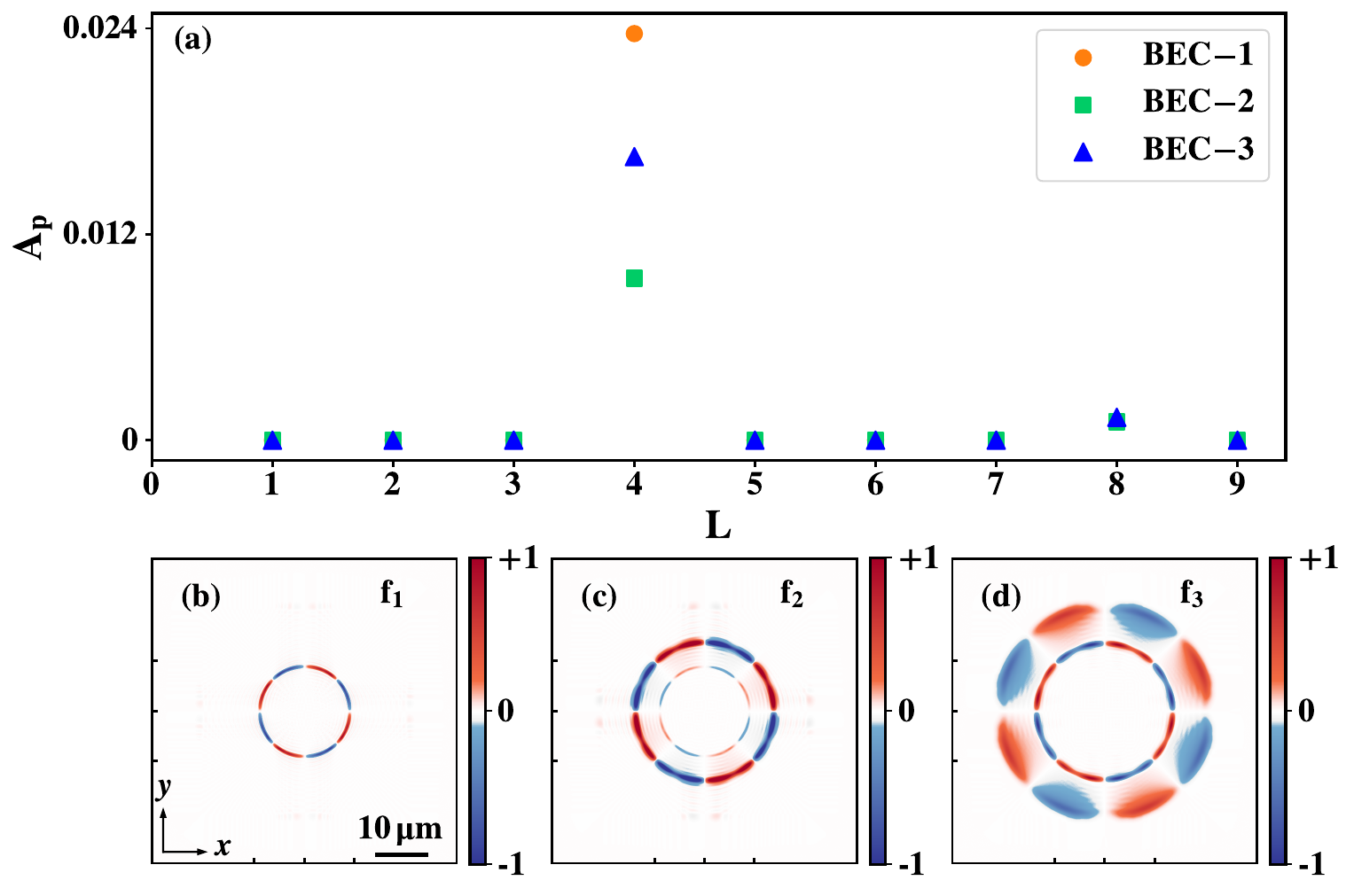}
    \caption{(a) The Normalized spectral weight profile in the real time dynamics for KHI. The most dominant Bogoliubov mode ($\mathrm{L=4}$) functions of BEC-1, BEC-2 and BEC-3 are shown in (b), (c), (d) respectively for KHI.}
    \label{fig:8}
\end{figure}

In the KHI regime, the maximum amplitude $\mathrm{A_p}$ is plotted in Fig.~\ref{fig:8}(a) for various low-lying collective excitation modes. It is observed that the fourth excited eigenmode provides the dominant contribution across all components. Figs.~\ref{fig:8}(b-d) show the spatial profiles of the corresponding Bogoliubov mode functions for this dominant mode ($\mathrm{L} = 4$). The mode exhibits a clear fourfold azimuthal structure, indicative of periodic density modulations along the interface~\cite{suzuki_crossover_2010}.

The selection of the dominant azimuthal mode number is governed by a competition between destabilizing and stabilizing contributions in the dispersion relation Eq.~\eqref{eqn::Frequency1}. We consider the interface deformation in the form $\eta_j \sim \sin(k_\theta \theta -\omega t)$, where $k_\theta$ is the dimensionless azimuthal mode number, such that $\mathrm{L}=k_\theta$ and taking the characteristic radial wavenumber as $k_r\sim1/r_j$. So The growth rate $\mathrm{Im}(\omega)$ is maximized at the mode $\mathrm{L}$ for which the destabilizing term $\rho_j\rho_{j+1}(\Delta\Omega^{\mathrm{eff}})^2k_\theta^2$ most strongly exceeds the stabilizing surface-tension term $\sigma_j k_r k_\theta^2/r_j^2 \sim \sigma_j \mathrm{L}^2/r_j^3$. Since the stabilizing term grows as $\mathrm{L}^2$ while the destabilizing term is independent of $\mathrm{L}$ at fixed $\Delta\Omega^{\mathrm{eff}}$, short-wavelength perturbations are suppressed by interfacial tension and quantum pressure. This produces a finite-wavelength instability with a preferred mode whose dominant angular symmetry depends on $\Delta\Omega^{\mathrm{eff}}$ and the initial seed. Numerical simulations with different phase-imprinted vorticities confirm this: a lower vorticity $l_2=8$ produces a twofold ($\mathrm{L}=2$) pattern, while $l_2=10$ and $l_2=20$ produce threefold ($\mathrm{L}=3$) and fourfold ($\mathrm{L}=4$) patterns, respectively, demonstrating that the dominant mode tracks $\Delta\Omega^{\mathrm{eff}}$ and is not fixed by the trap geometry.

\begin{figure}[t]
    \centering
    \includegraphics[width=1\linewidth]{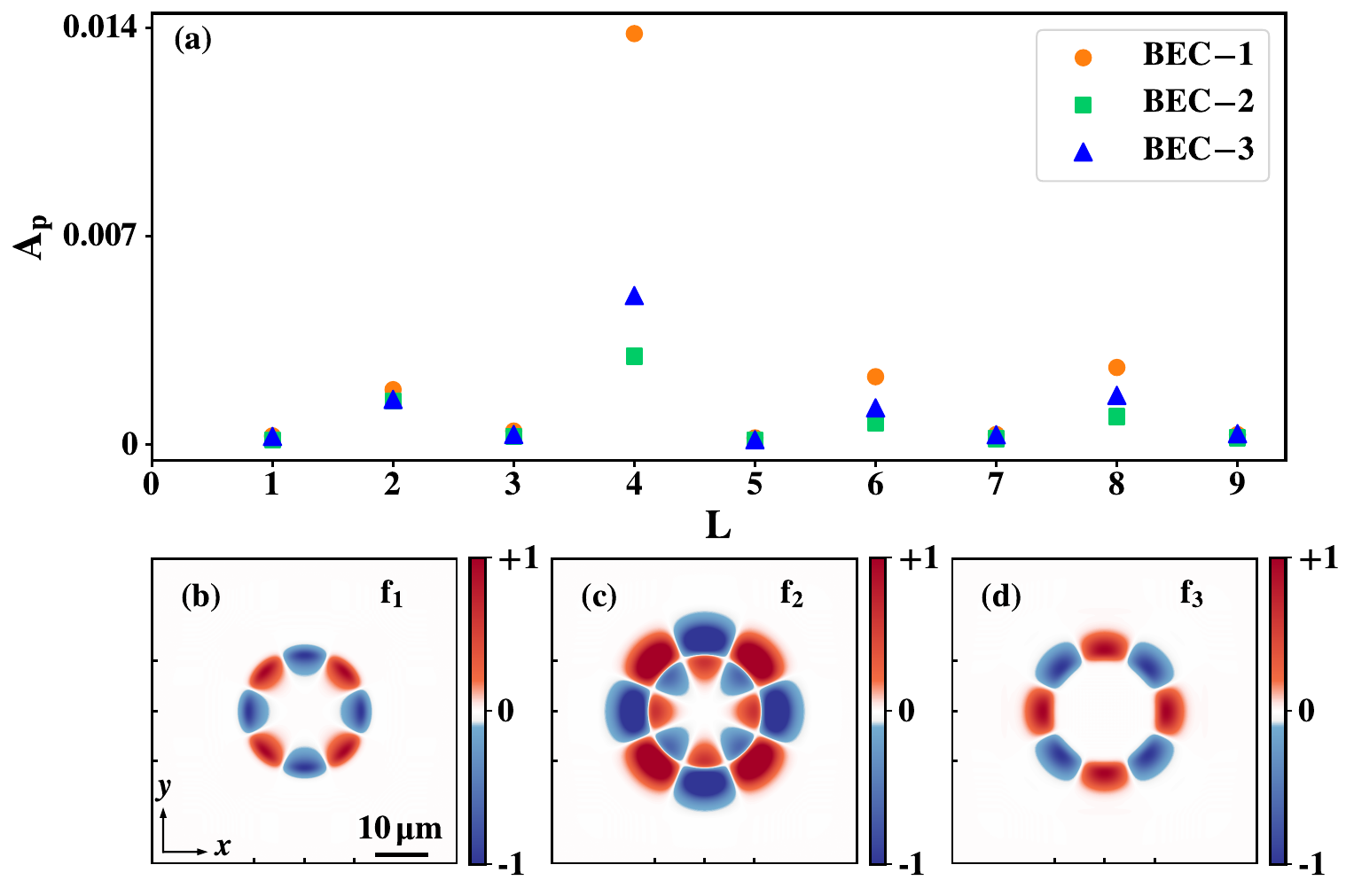}
    \caption{(a) The Normalized spectral weight profile in the real time dynamics for CSI. The most dominant Bogoliubov mode ($\mathrm{L=4}$) functions of BEC-1, BEC-2 and BEC-3 are shown in (b), (c), (d) respectively for CSI.}
    \label{fig:9}
\end{figure}

The Bogoliubov amplitudes are strongly localized near the sharp interfaces separating adjacent components, consistent with earlier theoretical predictions~\cite{PhysRevA.69.063608}. At each interface, the density fluctuations of the two neighboring components occur with opposite phase, such that regions of maximum fluctuation in one component coincide with minimum fluctuation in the other. The alternating red and blue lobes correspond to regions of opposite phase, forming an azimuthally modulated phase structure along the ring. In particular, BEC-1 and BEC-2 are out of phase at the inner interface, while BEC-2 and BEC-3 exhibit out-of-phase fluctuations at the outer interface.

As a result, the intermediate component (BEC-2), which is bounded by both interfaces, displays opposite fluctuation phases at its inner and outer boundaries. This behavior is characteristic of an interface-dominated collective excitation and clearly indicates that the KHI is driven primarily by surface modes localized along the interfacial regions.

For the CSI regime, the dominant Bogoliubov mode is also $\mathrm{L}=4$, as seen from the spectral weight $\mathrm{A_p}$ plotted in Fig.~\ref{fig:9}(a). Unlike the KHI case, however, all three components show comparable contributions at $\mathrm{L}=4$, reflecting that the instability is not interface-localized but involves the bulk of all overlapping components. In the same manner, Fig.~\ref{fig:9} presents the spatial profiles of the Bogoliubov mode functions with $\mathrm{L}=4$ in the CSI regime. In contrast to the KHI case, Fig.~\ref{fig:9} shows that the Bogoliubov amplitudes are distributed over a broader region across the thick and overlapping interface between adjacent components. This extended spatial distribution indicates that the instability is bulk in nature, originating from the relative counterflow within the overlapping region of the condensates. In this regime, significant compressible density fluctuations develop throughout the bulk, reflecting the compressible character of the excitation.

Overall, the stability analysis highlights the fundamental distinction between KHI and CSI: the former is dominated by surface-localized excitations confined to sharp interfaces, whereas the latter is governed by bulk excitations arising from compressible density modulations in strongly overlapping regions.

This contrast reflects the distinct physical origin and energy source of the two instabilities. The KHI mode is a surface wave drawing its free energy exclusively from the kinetic energy of interfacial shear flow; its amplitude is exponentially localized within a healing length of the sharp boundary, and the instability saturates through vortex nucleation confined to that boundary layer. The CSI mode is a compressible bulk excitation drawing its free energy from the relative counterflow of two interpenetrating superfluids throughout the finite-width overlap region; it saturates through vortex-antivortex pairs and soliton-like structures distributed across the overlap volume. In the three-component system, BEC-2 serves as a sensitive diagnostic: for KHI its Bogoliubov amplitude is confined to thin shells at its inner and outer edges with opposite oscillation phase at the two edges [Fig.~\ref{fig:8}(c)], whereas for CSI the amplitude spans the full radial extent of BEC-2 [Fig.~\ref{fig:9}(c)]. This spatial signature is in principle accessible via in-situ density imaging and could serve as an experimental diagnostic for distinguishing the two instability mechanisms.

\section{CONCLUSIONS}
\label{Conclusions} 
In the present paper, we have studied the generation and evolution of KHI, CSI, and their coexistence in a three-component BEC confined in a quasi-2D harmonic trap, using a unified framework based on differential effective rotation. The central physical insight is that selective central phase imprinting on individual components generates exact differential effective rotation $\Delta\Omega_j^{\mathrm{eff}} = (l_j - l_{j+1})\hbar/mr_j^2$ at each interface, the exact superfluid analogue of classical velocity shear. This follows directly and exactly from the phase ansatz $\phi_j = l_j\theta$ for a multiply-quantized central vortex. In the strongly immiscible regime, this vorticity contrast drives interface-localized KHI, characterized by interfacial modulations and vortex nucleation. In the weakly miscible regime, relative counterflow of bulk components drives CSI, manifested as extended density modulations. In the coexistence case, the scattering-length quench drives the system from a CSI-favorable to a KHI-favorable regime, while differential effective rotation $\Delta\Omega^{\mathrm{eff}}$ between components triggers the instabilities.
 
The BdG analysis establishes that the dominant unstable mode is selected by a competition between $\Delta\Omega^{\mathrm{eff}}$-driven destabilization and surface-tension stabilization; the dominant angular symmetry depends on $\Delta\Omega^{\mathrm{eff}}$ and the initial seed, with lower vorticity ($l_2=8$) selecting $\mathrm{L}=2$ and higher vorticity ($l_2=20$) selecting $\mathrm{L}=4$.

Our results demonstrate that a three-component BEC with two tunable interfaces exhibits instability dynamics qualitatively richer than those of binary mixtures in three specific ways. First, KHI and CSI can coexist simultaneously at different spatial locations in the same condensate.  Second, the shear at the two interfaces is independently controllable through the vorticity contrast $\Delta l_j$ at each boundary, giving two independent instability knobs absent from the binary case. Third, the intermediate component BEC-2, bounded by both interfaces, responds simultaneously to both instability modes and serves as a spectroscopic probe of both boundaries, a feature unique to the three-component geometry. The combination of selective central phase imprinting and tunable intercomponent interactions provides a clean, controllable, and experimentally accessible platform for studying multicomponent quantum hydrodynamics, extending the rich two-component phenomenology established in Refs.~\cite{takeuchi_quantum_2010, suzuki_crossover_2010} to a qualitatively new regime.

\section{ACKNOWLEDGEMENTS}

We acknowledge PARAM Shakti at the Indian Institute of Technology Kharagpur, a National Supercomputing Mission initiative by the Government of India, for providing the computational resources essential for this research. We gratefully acknowledge Subrata Das for his technical assistance. We thank Koushik Mukherjee for fruitful discussions and useful suggestions. S.G. acknowledges with gratitude the funding provided by the UGC Fellowship, India. A.S. and H.S.G. sincerely appreciate the support received through the Prime Minister’s Research Fellowship (PMRF), India. 
Vipin extends appreciation to the Ministry of Education(MHRD), Government of India, for the research fellowship.

\bibliographystyle{apsrev4-2}
\bibliography{References}

\end{document}